\documentclass[aps,prl,preprint,groupedaddress]{revtex4}
\usepackage{graphicx}
\begin{document}
\title{When hot water freezes before cold}
\author{J. I. Katz}
\email{katz@wuphys.wustl.edu}
\affiliation{Department of Physics and McDonnell Center for the Space Sciences\\Washington University, St.~Louis, Mo. 63130}
\date{\today}
\begin{abstract}
I suggest that the origin of the Mpemba effect (the freezing of hot water
before cold) is freezing-point depression by solutes, either gaseous or
solid, whose solubility decreases with increasing temperature so that they
are removed when water is heated.  They are concentrated ahead of the
freezing front by zone refining in water that has not been heated, reduce
the temperature of the freezing front, and thereby reduce the temperature
gradient and heat flux, slowing the progress of the freezing front.  I
present a simple calculation of this effect, and suggest experiments to test
this hypothesis.
\end{abstract}
\maketitle
\newpage

In a sub-freezing environment initially hot water often
freezes before initially cold water.  This observation is counter-intuitive
because one na\"\i vely expects the hot water first to cool to the
temperature of the initially cold water, and then to follow the cooling
history of the initially cold water.  However, the effect has been observed
many times and is folk-wisdom in many cultures; the earliest known reference
is by Aristotle.  It was brought to the attention of modern science by the
Tanzanian high school student for whom it is now named; with admirable
persistence in the face of disbelief on the part of his teachers, he
insisted on the primacy of empirical evidence over theory.
The history and literature are summarized by Auerbach \cite{A95}.

No generally accepted explanation of the Mpemba effect exists.  Apparently,
pre-heating water affects its properties in a manner that accelerates its
freezing.  A number of mechanisms have been considered, including loss
of mass by evaporation \cite{K69}, the loss of dissolved gases (whose
solubility in hot water is much less than in cold water) and supercooling
\cite{A95}.

Wojciechowski, {\it et al.\/} \cite{WOB88} report what appear to be the
only systematic quantitative measurements of the Mpemba effect.  Following
a suggestion of Freeman \cite{F79}, they also measured the freezing of
water saturated with CO$_2$.  Unfortunately, they did not describe the
ionic content of their water, which is likely to be essential.

Auerbach \cite{A95} found substantial (several degrees) but
non-reproducible supercooling in both preheated and non-preheated samples.
As a result, some of his data showed an Mpemba effect while some did not.
His experiments were performed on distilled and de-gassed
water, and may therefore not be applicable to observations of the Mpemba
effect in tap water or environmental water (Mpemba's observations were on
sugared milk he was freezing to make ice cream!).


The observations clearly point to some change in water when heated.  As
has been remarked before, heating water removes dissolved air (chiefly
nitrogen) because its solubility decreases rapidly with increasing
temperature.  The problem is to find a mechanism by which the removal
of a small quantity of dissolved material (the solubility of nitrogen in
water at room temperature is only about 0.7 mmolar/bar) can produce
a Mpemba effect.

Gases are not the only substances whose solubility in water decreases with
increasing temperature.  Most natural waters are ``hard'', containing a
variety of dissolved mineral salts, most importantly calcium bicarbonate
Ca(HCO$_3$)$_2$.  This is introduced into ground water that has been 
acidified by atmospheric carbon dioxide by the reaction in limestone
rock \cite{P55}
\begin{equation}
\rm CaCO_3 + CO_2 + H_2O \leftrightarrow Ca(HCO_3)_2.
\end{equation}
Because the solubility of gases in liquids decreases rapidly with increasing
temperature, heating shifts the equilibrium to the left, resulting in the
precipitation of limestone deposits known as ``boiler scale'' or ``kettle
fur''.  For this reason, hardness resulting from bicarbonates is known as
``temporary hardness'' \cite{P55}.  Hard water that has been heated loses
much of its dissolved calcium.  Hence the freezing point $T_m^\prime$ of
never-heated hard water is lower, because of the depression of freezing
points by solutes \cite{G67}, than the freezing point $T_m$ of hard water
that has been heated.  For the same thickness of the ice layer, the
temperature gradient and heat flux are less in the never-heated water,
so it takes longer to lose the latent heat of freezing and freezes more
slowly.  This is illustrated in Fig. 1.
\begin{figure}
\includegraphics[width=0.7\textwidth]{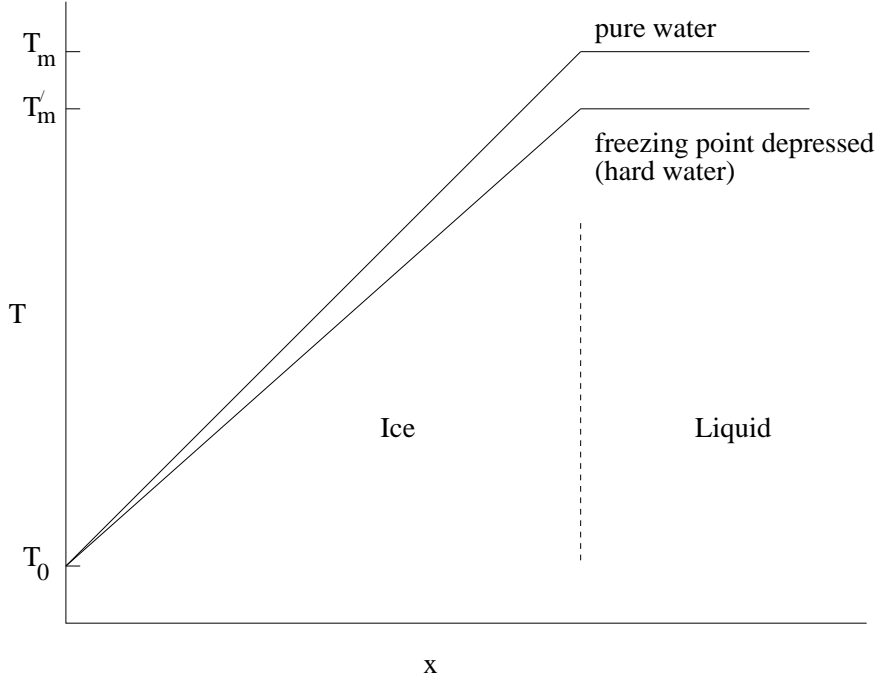}
\caption{\label{figure1} Temperature profiles, neglecting specific heat in
comparison to latent heat.  $T_m$ is the freezing point of pure water,
$T_m^\prime < T_m$ that of hard water, lowered by freezing point depression.
Its temperature gradient and heat flux are reduced, so the freezing front
advances more slowly than in water purified by heating.}
\end{figure}

The concentration of dissolved calcium in moderately hard tap water is about
100 ppm, or 2.5 mmolar, but values several times greater or less are found.
Each Ca$^{++}$ ion corresponds to a total of three dissolved ions, so such
a solution has a total ion concentration of 7.5 mmolar.  The freezing point
depression of ideal dilute aqueous solutions \cite{G67} is $\Delta T_m =
R T_m^2/H_m = 1.86^\circ\,$C/molar, where $R$ is the gas constant and $H_m$
the enthalpy of melting.  The effect of hardness is typically ten times
greater than that of dissolved air, but still rather small without further
concentration.

The equilibrium ratio of the concentration of a solute in a solid to that in
its melt is the Nernst equilibrium distribution coefficient $k_0$ \cite{G67}.
For most solutes in water and ice $k_0 \ll 1$; they are almost completely
excluded from the solid.  As a freezing front advances into water the
solutes are pushed ahead of it, and their concentration close to the front
is enhanced over its initial value in the liquid, a process known as zone
refining \cite{P66}.

The conventional theory of zone refining \cite{BPS53} assumes a stationaryi
state with a boundary condition that the concentration of solute equals its
initial concentration in the liquid at a finite distance from the freezing
front.  This latter condition is appropriate when there is driven fluid
circulation outside a viscous boundary layer of thickness $\delta$.  In the
present problem there is no circulation and $\delta \to \infty$.
Worse, in the limits $k_0 \to 0$ and $\delta \to \infty$ this solution
becomes singular because in steady state the solute from an infinite volume
of fluid must accumulate in a finite region of enhanced concentration.

Our problem corresponds to the low concentration limit of the theory
of the freezing of binary melts \cite{HW85}, in which the progress
of the freezing front is controlled by diffusion of heat.  Mathematically,
it is a ``Stefan problem'' in which the boundary condition is moving.  We
neglect the specific heats of the liquid and solid in comparison to the
latent heat of the phase transition.  Then the liquid is isothermal and
there is no thermal diffusion of the solute within it.

The freezing front advances at a nonsteady speed $v_f(t) = \sqrt{(T_m - T_0)
\kappa / (2 H_m \rho_i t)}$ \cite{S1891}, where $T_m$ is the melting
temperature, $T_0 < T_m$ is the subfreezing boundary temperature, $\kappa$
is the thermal conductivity and $\rho_i$ the density of ice, and $t$ is the
time elapsed since freezing began.  The equation for the conservation of
solute concentration $C(x,t)$ in the liquid, in a frame moving with the
freezing front, is
\begin{equation}
{\partial C(x,t) \over \partial t} + v(t) {\partial C(x,t) \over \partial x}
- D {\partial^2 C(x,t) \over \partial x^2} = 0,
\end{equation}
where $v(t) = - v_f(t) \rho_i/\rho_l$ is the velocity at which liquid flows
to the freezing front and $D$ is the diffusion coefficient of solute in the
liquid.

Both the diffusion length $\sqrt{Dt}$ and the frozen layer thickness $\int
v_f(t)\,dt$ are proportional to $t^{1/2}$.  Hence the solution is stationary
in suitable variables.  Defining $\zeta \equiv x/\sqrt{Dt}$ \cite{LC1831},
we obtain the separable equation
\begin{equation}
t {\partial C(\zeta,t) \over \partial t} - (K_M + {\zeta \over 2}) {\partial
C(\zeta,t) \over \partial \zeta} - {\partial^2 C(\zeta,t) \over \partial
\zeta^2} = 0,
\end{equation}
where the dimensionless parameter $K_M \equiv \sqrt{(T_m - T_0) \kappa
\rho_i / (2 H_m \rho_l^2 D)}$.  The value of $D$ for Ca(HCO$_3$)$_2$ in
water is not readily available, so we take the values \cite{AIP72,HPQ97} for
CaCl$_2$ as estimates and extrapolate to 0$^{\circ\,}$C, giving $D \approx
0.6 \times 10^{-5}$ cm$^2$/sec.  Then $K_M \approx 7 \sqrt{(T_m - T_0)/
10^{\circ\,} {\rm C}}$.

Writing $C(\zeta,t) = C_\zeta(\zeta) C_t(t)$, we find ordinary differential
equations for $C_\zeta(\zeta)$ and $C_t(t)$ with a separation constant $A$:
\begin{eqnarray}
{d\ln C_t(t) \over d\ln t}&= A\\
{d^2 C_\zeta(\zeta) \over d\zeta^2} + (K_M + {\zeta \over 2})
{dC_\zeta(\zeta) \over d\zeta} - A C_\zeta(\zeta)&= 0.
\end{eqnarray}
For $K_M \gg 1$ and $\zeta = {\cal O}(1)$ we can neglect the $\zeta/2$ term
and the solution is elementary.  From the conservation of solute (and $k_0
\to 0$) we find $A = 0$.  The equation for $C_\zeta(\zeta)$ has solutions
$C_\zeta(\zeta) \propto \exp{(\gamma\zeta)}$ with $\gamma =$ 0, $-K_M$.  The
root $\gamma = 0$ gives the uniform solute density far from the freezing
front, while the root $\gamma = - K_M$
gives the enhanced concentration of solute close to the front:
\begin{equation}
C_\zeta(\zeta) = C_0 [1 + 2 K_M^2 {\rho_l \over \rho_i} \exp{(-K_M \zeta)}].
\end{equation}
This justifies the neglect of the $\zeta/2$ term to ${\cal O}(K_M^{-2})$.

The concentration at the freezing front is enhanced by a factor $\approx
100 (T_m -T_0)/10^{\circ\,}{\rm C}$.  The fractional freezing point
depression
\begin{equation}
{\Delta T \over T_m - T_0} = {R T_m^2 \kappa \over H_m^2 \rho_l D} C_0,
\end{equation}
where $C_0$ is the concentration in moles/cm$^3$, independent of time and of
$T_0$.  For Ca(HCO$_3$)$_2$ in water the coefficient of $C_0$ is 20/molar
so an ion concentration of 5 mmolar (67 ppm calcium) produces a 10\%
reduction in freezing rates compared to pure (once heated) water.

Wojciechowski, {\it et al.\/} \cite{WOB88} reported an apparent increase of
$H_m$ of CO$_2$-saturated water of $10 \pm 4$ cal/gm.  This is not likely to
be an actual change in $H_m$, but may reflect a reduction of the freezing
point at the freezing front of $10 \pm 4^{\,\circ}$C, requiring the removal
of additional internal energy before freezing begins.  Application of the
theory presented here predicts that for water saturated with respect to
CO$_2$ at 1 bar pressure at room temperature (33 mmolar CO$_2$) and $T_m -
T_0 = 23^{\,\circ}$C \cite{WOB88} the freezing point depression at the front
is $10^{\,\circ}$C, consistent with the apparent increase in $H_m$.  These
authors found an Mpemba effect only in water that had not been degassed by
boiling, which is consistent with the removal of temporary hardness, even
though dissolved air itself is predicted to produce only a small Mpemba
effect.

Our model makes readily testable predictions for the dependence of a
Mpemba effect on the concentration of solutes whose solubility decreases
with increasing temperature.  It also predicts no Mpemba effect, but a
dependence of freezing time on concentration, for solutes whose solubility
does not decrease with increasing temperature.  It predicts that the
magnitude of freezing point depression and of the effect (having subtracted
the time required for the solutions to cool to the freezing point) should
be independent of the thickness of ice formed.  In this model an Mpemba
effect occurs in the time required for formation of an ice layer of any
thickness, and hence is found for the onset of freezing as well as for
complete solidification.  Fully quantitative predictions require modeling of
heat transfer through the air boundary layer.

I thank Nadya MacAloon, a student in my ``How Things Work'' class, for
asking if the Mpemba effect is real, and L. M. Canel for finding and
correcting an error.


\begin{thebibliography}{99}
\bibitem{A95} D.~Auerbach, ``Supercooling and the Mpemba effect: When hot
water freezes quicker than cold,'' Am.~J.~Phys. {\bf 63}, 882--885 (1995).

\bibitem{K69} G.~S.~Kell, ``The freezing of hot and cold water,''
Am.~J.~Phys. {\bf 37}, 564 (1969).

\bibitem{WOB88} B.~Wojciechowski, I.~Owczarek and G.~Bednarz, ``Freezing
of aqueous solutions containing gases,'' Crystal Res.~Tech. {\bf 23}, 
843--848 (1988).

\bibitem{F79} M. Freeman, ``Cooler still---an answer?'' Phys.~Educ. {\bf
14}, 417--421 (1979).

\bibitem{P55} L.~Pauling {\it College Chemistry\/} 2nd ed., (W. H. Freeman,
San Francisco 1955).

\bibitem{G67} E.~A.~Guggenheim {\it Thermodynamics\/} 5th ed., (North
Holland, Amsterdam 1967).

\bibitem{P66} W.~G.~Pfann {\it Zone Melting\/} 2nd ed., (Wiley, New York
1966).

\bibitem{BPS53} J.~A.~Burton, R.~C.~Prim and W.~P.~Slichter, ``The
Distribution of Solute in Crystals Grown from the
Melt.~Part I.~Theoretical,'' J.~Chem.~Phys. {\bf 21}, 1987--1991 (1953).

\bibitem{HW85} H.~E.~Huppert and M.~G.~Worster, ``Dynamic solidification of
a binary melt,'' Nature {\bf 314}, 703--707 (1985).

\bibitem{S1891} J.~Stefan, ``\"Uber die Theorie der Eisbildung, insbesondere
\"uber die Eisbildung im Polarmeere,'' Annalen der Physik und Chemie {\bf
42}, 269--286 (1891).

\bibitem{LC1831} G.~Lam\'e and B.~P.~Clapeyron, ``M\'emoire sur la
solidification par refoidissement d'un globe solide,'' Ann.~Chem.~Phys. {\bf
47}, 250--256 (1831).

\bibitem{AIP72} L.~G.~Longsworth, in {\it American Institute of Physics
Handbook\/} 3rd ed., ed. D.~E.~Gray (McGraw-Hill, New York, 1972) p.~2-222.

\bibitem{HPQ97} A.~V.~Eletskii, in {\it Handbook of Physical Quantities\/}
eds.  I.~S.~Grigoriev and E.~Z.~Meilikhov (CRC Press, Boca Raton, Fla.,
1997) p.~473.
\end{thebibliography}
\end{document}